\documentclass[aps,preprint]{revtex4}
\usepackage{epsfig}
\usepackage{graphicx}

\begin{document}

\title{Signature of effective mass in crackling noise asymmetry}

\author{Stefano Zapperi}
\affiliation{INFM unit\`a di Roma 1 and SMC, Dipartimento di
Fisica, Universit\`a "La Sapienza", P.le A. Moro 2, 00185 Roma, Italy}
\affiliation{Istituto dei Sistemi Complessi,
CNR, Via dei Taurini 9, 00185 Roma, Italy}
\author{Claudio Castellano}
\email[Correspondence should be addressed to CC.
Electronic address: ]{castella@pil.phys.uniroma1.it}
\affiliation{INFM unit\`a di Roma 1 and SMC, Dipartimento di
Fisica, Universit\`a "La Sapienza", P.le A. Moro 2, 00185 Roma, Italy}
\affiliation{Istituto dei Sistemi Complessi,
CNR, Via dei Taurini 9, 00185 Roma, Italy}
\author{Francesca Colaiori}
\affiliation{INFM unit\`a di Roma 1 and SMC, Dipartimento di
Fisica, Universit\`a "La Sapienza", P.le A. Moro 2, 00185 Roma, Italy}
\affiliation{Istituto dei Sistemi Complessi,
CNR, Via dei Taurini 9, 00185 Roma, Italy}
\author{Gianfranco Durin}
\affiliation{IEN Galileo Ferraris, str. delle Cacce 91, 10137 Torino, Italy}

\maketitle

\noindent{\bf Crackling noise is a common feature in many dynamic
systems
\cite{SET-01,DUR-04,FIE-95,COL-02a,MIT-01,PET-94,MIG-01,HOU-98,MET-04},
the most familiar instance of which is the sound made by a sheet of
paper when crumpled into a ball. Although seemingly random, this noise
contains fundamental information about the properties of the system in
which it occurs. One potential source of such information lies in the
asymmetric shape of noise pulses emitted by a diverse range of noisy
systems \cite{MEH-02,SPA-96,DUR-02,HOU-98,MET-04}, but the cause of
this asymmetry has lacked explanation \cite{SET-01}.  Here we show
that the leftward asymmetry observed in the Barkhausen effect
\cite{DUR-04} - the noise generated by the jerky motion of domain
walls as they interact with impurities in a soft magnet - is a direct
consequence of a magnetic domain wall's negative effective mass. As
well as providing a means of determining domain wall effective mass
from a magnet's Barkhausen noise our work suggests an inertial
explanation for the origin of avalanche asymmetries in crackling noise
phenomena more generally.}

Crackling noise is the response of many physical systems to a slow
external driving: outbursts of activity (avalanches or pulses)
spanning a broad range of sizes, separated by quiescent
intervals~\cite{SET-01}.  In condensed matter, notable examples are
the magnetization noise emitted along the hysteresis loop in
ferromagnets (i.e., the Barkhausen effect \cite{DUR-04}), the noise
from magnetic vortices in type II superconductors \cite{FIE-95},
ferroelectric materials \cite{COL-02a} and driven ionic crystals
\cite{MIT-01}. In the context of mechanics, examples are the acoustic
emission signal in fracture \cite{PET-94} and plasticity \cite{MIG-01}
and, on a larger scale, seismic activity in correspondence to an
earthquake \cite{HOU-98,MET-04}.  Quantitative understanding of
crackling noise is of fundamental importance in different
applications, from non-destruc\-tive material testing to hazard
prediction. This goal can be achieved only through the identification
of general universal properties common to these systems, irrespective
of their differences in the internal dynamics and microstructural
details. In this context, the average shape of the individual pulses
composing the signal has been recently proposed as the best tool to
characterize these universal features of crackling noise
\cite{SET-01}.  In analogy with critical phenomena, it is expected
that pulses of different durations can be rescaled on a universal
function, whose shape would only depend on general features of the
physical process underlying the noise. This scenario is supported by
the analysis of a variety of models, where pulse shapes are described
by universal symmetric scaling functions \cite{MEH-02,OBR-94,COL-04a}. In
most experimental data, however, the pulse shape is markedly
asymmetric with respect to its midpoint, i.e. avalanches start fast
but return to zero more
slowly \cite{SET-01,MEH-02,SPA-96,DUR-02,HOU-98,MET-04}. These
results are puzzling because the models accurately reproduce several
other universal quantities, such as avalanche distributions and power
spectra \cite{DUR-02,DUR-00}.

One of the most studied examples of crackling noise is the Barkhausen
effect recorded in soft magnetic materials. The Barkhausen noise is
due to the motion of domain walls, the interfaces separating regions
of opposite magnetization, in response to a magnetic field 
(see the movie as Supplementary Information).  Domain
walls are characterized by an effective mass \cite{DOR-48,Hubert},
which is related to the increase of the wall energy with velocity,
as experimentally revealed, for instance, in the dynamic
susceptibility of insulating ferrite materials \cite{RAD-50}. In
metallic ferromagnets, however, inertial effects
are usually neglected, being much smaller than eddy current
dissipation \cite{Bertotti}. This approximation is
usually assumed in the description of the Barkhausen effect 
\cite{DUR-04}.

We show in this Letter that the asymmetry of pulse shapes is a direct
signature of the effective mass associated to the objects moving under
the action of the external field.  In conducting ferromagnets the mass
is {\it negative} and this results in a leftward asymmetry of the
Barkhausen noise pulses. Our findings clarify the general meaning of
pulse shapes in crackling noise phenomena: the form of avalanche
displays an asymmetry that depends on the duration, and encodes
important information on the characteristic time of the underlying
dynamics. Only on very long timescales pulse shape symmetry 
and universality are recovered.

\begin{figure}

    \includegraphics[width=9cm,angle=-90]{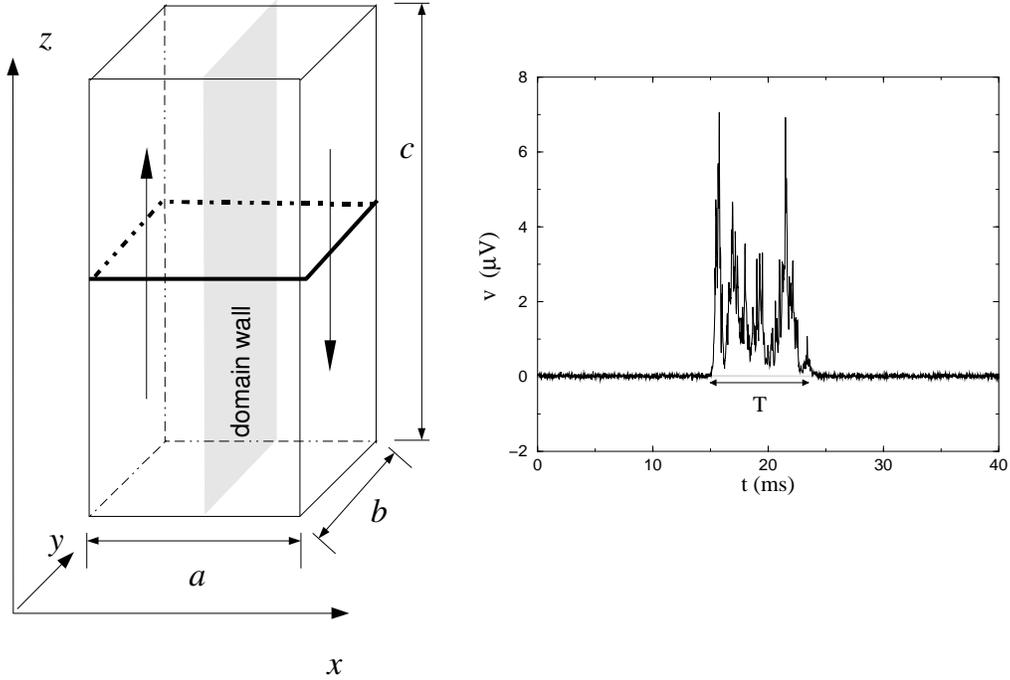}

    \caption {\label{fig:1} {\bf Schematic picture of the experimental
    setup.}  A ribbon of dimensions $a \times b \times c$ is placed in
    a solenoid (not shown) yielding a magnetic field in the vertical
    direction. The magnetization reversal process occurs via the displacement
    of domain walls, such as the one depicted here. The arrows indicate
    the directions of the magnetization. A pickup coil, wound
    around the sample, records a voltage signal $v$ that is composed by
    distinct pulses, as the one depicted in the right panel.}

\end{figure}

\begin{figure}
 
\includegraphics[width=12cm]{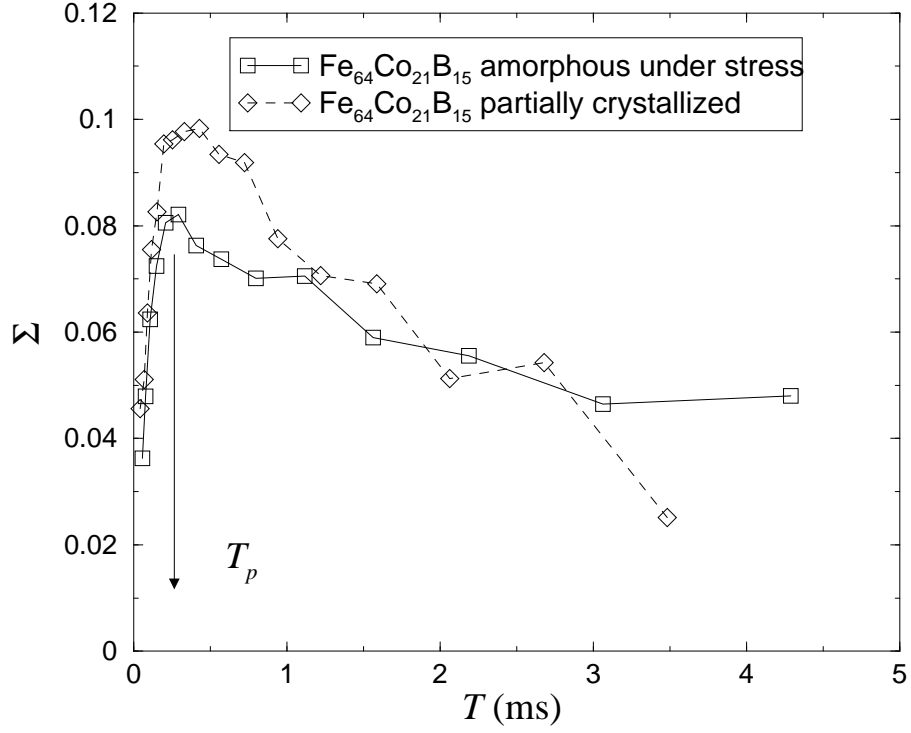}
\caption {\label{fig:4} {\bf Skewness of avalanches in experiments.}
The skewness of the pulse shape as a
function of the pulse duration obtained from experiments. The data
display a peak at $T_p^{exp} \simeq 200\mu s$ indicated by an arrow.}

\end{figure}

We measure Barkhausen noise pulse shapes in two ferromagnetic
alloys. The experimental setup is depicted in Fig.~\ref{fig:1}
(see Methods section).  In response to an increasing
external field, domain walls move, inducing in the pickup coil a
voltage signal $v(t)$ proportional to the magnetization rate. The
shape of the pulse is defined as the voltage $\langle v(t,T)\rangle$
at time $t$ averaged over all avalanches of duration $T$.  When
appropriately normalized and plotted as a function of $t/T$, the
experimentally measured shapes for different durations rescale fairly
well (though not perfectly, see below) on a universal function (see
Fig.~\ref{fig:2}, left), that is clearly asymmetric, in agreement with
earlier measurements \cite{MEH-02,SPA-96,DUR-02}.  Such a leftward
asymmetry implies that pulses start rapidly and decay slowly. This is
precisely the opposite of the effect of standard inertia, the resistance
of a body to changes on its motion, which would imply a slow increase of the
velocity when the domain wall is at rest.  The asymmetry can be
quantified by computing the average skewness
\begin{equation}
\Sigma(T) =
{
{1 \over T} \int_0^T dt \langle v(t,T) \rangle
\left( t- \bar t \right)^3
\over
\left[{1 \over T} \int_0^T dt \langle v(t,T) \rangle
\left( t- \bar t \right)^2
\right]^{3/2}
}
\end{equation}
where $\bar t = 1/T \int_0^T dt \langle v(t,T) \rangle t$.
As shown in Fig.~\ref{fig:4}, in both samples the skewness
is always positive, indicating a leftward asymmetry, and it displays
a peak for $T_p^{exp} \approx 200\mu s$.

To account for these experimental results, we make use of a simple and
successful model for the Barkhausen effect in soft metallic ferromagnets,
based on the dynamics of a planar domain wall in an effective pinning
potential \cite{ALE-90}.  The equation of motion for the wall position
$x$ is given by
\begin{equation}
\beta\dot{x}= 2I_s(c_H t-k x +W(x)), \label{eq:abbm}
\end{equation}
where $\beta$ is a damping constant, $I_s$ is the saturation
magnetization, $c_H t$ is the external field increasing at rate $c_H$,
$-k x$ is the demagnetizing field and $W(x)$ is a random field with
Gaussian distribution and Brownian correlations \cite{ALE-90}. 
These correlations are believed to represent an effective description
of a more general model with flexible domain walls \cite{ZAP-98,MIL-02}.
Eq.~(\ref{eq:abbm}) can be solved exactly and provides an excellent
description of the statistical properties of the Barkhausen noise,
considering that the domain wall velocity $\dot{x}$ is proportional to
the recorded voltage signal $v(t)$. The solution of the model,
however, yields a symmetric pulse shape \cite{OBR-94,COL-04a}, at odds
with experimental evidence.

\begin{figure}
 
\includegraphics[width=12cm]{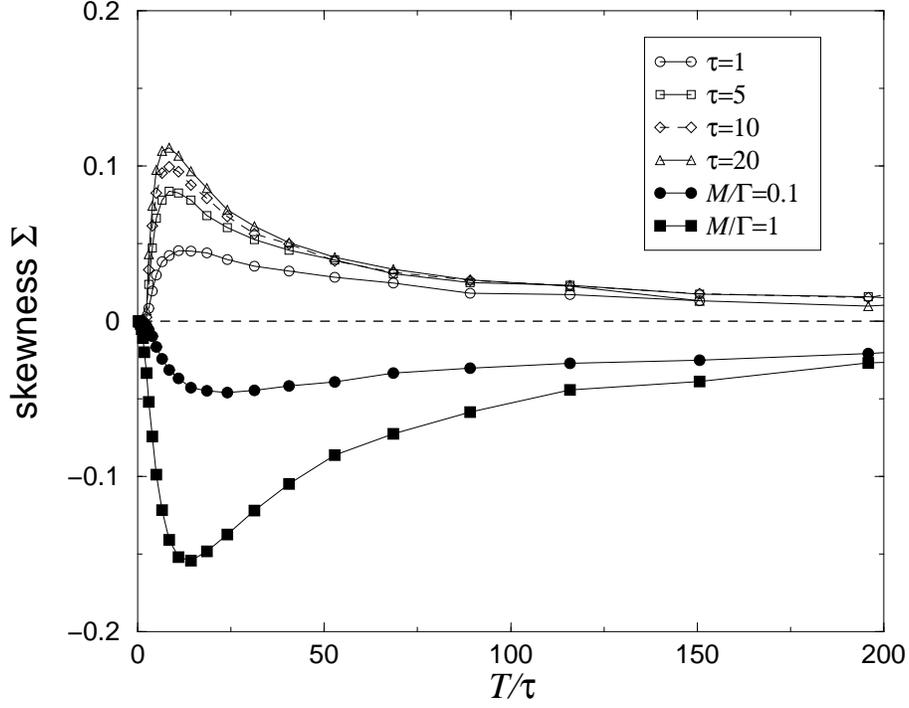}

\caption {\label{fig:3} {\bf Skewness of avalanches in the model.}
The skewness of the pulse shape obtained in the domain wall model
for different values of eddy current time constant $\tau$ is plotted
as a function of the pulse duration $T$. We also report
a similar measurement in the case of conventional positive domain wall
mass. In the latter case the skewness is negative (rightward asymmetry).}

\end{figure}

The assumption used to derive Eq.~(\ref{eq:abbm}) is that at each time
the work done by the effective field (i.e. the applied field corrected
by the demagnetizing and pinning fields) is compensated by the energy
dissipated by eddy currents, which is estimated in the quasistatic
approximation \cite{ALE-90}. A more detailed analysis of eddy current
dissipation, including dynamic effects (see the Methods section
and Ref.~\cite{BIS-80}), leads to the identification of a
frequency-dependent effective mass, which turns out to be negative in
the entire spectrum and equal to $M^* \approx - \beta \tau/(2 \gamma)$
at low frequencies, where $\tau$ is the longest relaxation time and
$\gamma \approx 1.05$.  For the materials considered in our
experiments, the effective mass can be estimated to be $M^* \approx -7\cdot
10^{-5} kg/m^2$, much larger than the positive D\"oring  domain
wall mass ($M\approx 10^{-9} kg/m^2$ \cite{DOR-48}).  This negative
inertial effect can also be formulated by adding a non-local damping term
in the equation of motion for the wall (see Methods section)
\begin{equation}
\Gamma\dot{x}+{\Gamma_0 \over \tau} \int^t e^{-(t-t')/\tau} \dot{x}(t')dt'
= 2I_s(c_H t-k x +W(x)), \label{eq:abbm2}
\end{equation}
where $\Gamma$ and $\Gamma_0$ are coefficients of the same
order of magnitude with $\Gamma+\Gamma_0=\beta$.
In order to understand the role of the effective mass on the pulse
shape, we have numerically integrated Eq.~(\ref{eq:abbm2}) for
different values of $\tau$. 
While the distributions of avalanches duration and size are unaffected
by the addition of the inertial term, the pulse shapes become asymmetric
and bear a remarkable similarity with the experimental ones
(see Fig.~\ref{fig:2}, right).
Also in this case the skewness is always positive (Fig.~\ref{fig:3}),
indicating a leftward asymmetry, and it displays a peak in correspondence
with a characteristic time $T_p \approx 10 \tau$.

Using the value of the largest relaxation timescale for the samples
 considered experimentally (i.e. $\tau=\mu\sigma b^2/\pi^2 \approx 5
 \mu s$, where $b$ is the sample thickness, $\mu$ is the permeability
 and $\sigma$ the conductivity), we obtain a theoretical estimate of
 the peak position $T_p \approx 50 \mu s$, which is reasonably close
 to the value measured experimentally,
 considering  the approximations involved in the model.
 In particular, we have treated a single
 domain wall moving at the center of the sample, while in the
 experiments several domain walls are present. This induces a non
 trivial interference among the eddy current patterns generated by
 each wall.  To corroborate our interpretation of the asymmetry in
 terms of a negative effective mass, we have performed simulations of
 Eq.~(\ref{eq:abbm2}) replacing the non-local term with a positive
 mass term $M \ddot{x}$~\cite{BAL-03}. The resulting skewness, reported in
 Fig.~\ref{fig:3}, is negative and displays a peak at $T_p \approx 10
 M/\Gamma$.

In conclusion, we have shown that the pulse shape asymmetry commonly
observed in Barkhausen noise measurements in metals is a signature of
a negative effective mass of domain walls. Remarkably, the skewness
exhibits a peak that can be used to track the characteristic
relaxation timescale, corresponding to the ratio between mass and
damping constant. Similar asymmetric pulse shapes are also observed
in seismic data, recorded in correspondence to an earthquake
\cite{HOU-98,MET-04}. Seismic movements are due to the motion of fault
planes in response to the stress accumulated in the crust. A non-local
dynamical effect, analogous to the one discussed here, could be due to
the presence of stress overshoots \cite{SCH-01a}. It would be
interesting to test this mechanism using existing fault dynamics
models.

\newpage

\section*{Methods}

\subsection*{Materials and experimental methods}

In our experiments a long solenoid provides an homogeneous driving
field ${\bf H}$ ramped at constant rate  $c_H=4 A/(m\;s)$ , while a
secondary pickup
coil around the sample cross section measures the induced flux. The
pickup coil is made of 50 isolated copper turns, wound within 1
mm. Such a small width is required to avoid spurious effects due to
demagnetizing fields. The measurements are performed only in the
central part of the hysteresis loop around the coercive field, where
domain wall motion is the relevant magnetization mechanism {\it (1)}.
In this case the recorded voltage signal $v(t)$ is
proportional to the domain wall velocity $\dot{x}$. To reduce excess
external noise during the measurement, we use a low pass
pre-amplifier filter with a cutoff frequency of 100 kHz.  We employ
two ribbons of Fe$_{64}$Co$_{21}$B$_{15}$, having width $a=1 cm$,
thickness $b=24 \mu m$, and length $c=20 cm$: the first sample is
amorphous, and subjected to a small tensile stress (2MPa). The second has
been partially crystallized (5\%) by thermal annealing. Tensile
stress and thermal annealing are used to relax part of the internal
stresses, leading to a high signal to noise ratio.

\subsection*{Derivation of domain wall motion from eddy current dissipation}

Considering the configuration described in Fig.~\ref{fig:1}, the
relevant component of field ${\bf H_e}=H_e(x,y,t)\hat{z}$, associated
with the eddy currents produced by the moving domain wall, obeys the
equation
\begin{equation}
\sigma\mu \frac{\partial H_e}{\partial t} -\nabla^2 H_e=0,
\label{eq:eddy}
\end{equation}
where  $\sigma$ is the conductivity and $\mu$ is the permeability
\cite{Bertotti}. One should complement Eq.~(\ref{eq:eddy}) with the Faraday
condition around the wall
\begin{equation}
\partial_x  H_e(0^+,y,t)-\partial_x H_e (0^-,y,t) = 2\sigma I_s v_x(t),
\label{eq:fara1}
\end{equation}
where $v_x(t)$ is the domain wall velocity and we have assumed $c=\infty$.
Eqs.~(\ref{eq:eddy}) and
(\ref{eq:fara1}) are solved with the boundary condition $H_e=0$ at
the sample surface. The mean pressure on the wall is obtained as
\begin{equation}
P_e={2I_s \over b} \int_{-b/2}^{b/2} H_e(0,y,t) dy = \int dt' f(t-t') v_x
(t'). \label{eq:pe}
\end{equation}
A direct calculation of the response function, similar to Ref.~\cite{BIS-80},
leads to
\begin{equation}
f(t)=-\frac{32 I_s^2}{\mu a \pi^2} \theta_2[e^{-4 \pi^2 t/(a^2 \mu \sigma)}]
\sum_{n=0}^{\infty} \frac{e^{-t/\tau_n}}{(2n+1)^2},
\label{eq:response}
\end{equation}
where $\theta_2$ is the Jacobi Elliptic Function and
the relaxation times are $\tau_n = \mu \sigma b^2/[(2n+1)\pi]^2$.

The equation of motion for the domain wall is obtained by
equating the eddy current pressure $P_e$ to the pressure $P_a= 2I_s
H$ exerted by the applied field corrected by demagnetizing effects
and pinning. The left-hand side of Eq.~(\ref{eq:abbm2}) is recovered
by considering the leading contributions at short and long times, involving
the largest relaxation time $\tau \equiv \tau_0=\mu \sigma b^2/\pi^2$.
The constants can be estimated as
\begin{equation}
\Gamma = {16 I_s^2 \sigma b \over \pi^3}
\left( \gamma - {2\alpha \over \pi} \right) ~~~~~~
 \Gamma_0  = {32 I_s^2 \alpha \sigma b \over \pi^4}
\end{equation}
with $\alpha \approx 0.886$ and
$\gamma = \sum_{n=0}^\infty 1/(2 n+1)^3 \approx 1.05$.

In the frequency domain, we can formally rewrite Eq.~(\ref{eq:pe})
as $\tilde{P}_e=(\beta + i \omega M^*) \tilde{v}_x$, with
the effective mass, in the low frequency limit, given by
\begin{equation}
M^*\approx - {8 I_s^2 b^3 \mu \sigma^2  \over \pi^5 }
 = -{\beta \tau \over 2 \gamma}.
\end{equation}

\section*{Statement on financial interests}

The authors declare no competing financial interest.

\newpage

\begin{figure}

   \includegraphics[width=14cm]{Figure2.eps}

   \caption {\label{fig:2} {\bf Comparison of the average pulse shapes in
the experiments and in the model.} On the left panel we report the average
pulse shape obtained from Barkhausen noise measurements in a partially
crystallized Fe$_{64}$Co$_{21}$B$_{15}$ ribbon. The shapes for
different durations $T$ are normalized and rescale quite well, apart
from a small systematic variation related to the asymmetry.  On the right
panel, we report the corresponding shapes obtained from the model.
The normalization constant is given by
$N=\int_0^T dt \langle v(t,T)\rangle/T$.}

\end{figure}

\end{document}